\documentclass[10pt, a4paper, twocolumn]{article} 
\usepackage[english]{babel} % English language hyphenation
\usepackage{amsmath,amsfonts,amsthm,amssymb} % Math packages for equations
\usepackage[svgnames]{xcolor} % Enabling colors by their 'svgnames'
\usepackage[hang, small, labelfont=bf, up, textfont=it]{caption} % Custom captions under/above tables and figures
\usepackage{booktabs} % Horizontal rules in tables
\usepackage{graphicx} % Required for adding images
\usepackage{enumitem} % Required for customising lists
\setlist{noitemsep} % Remove spacing between bullet/numbered list elements
\usepackage{sectsty} % Enables custom section titles
\usepackage{parskip}
\usepackage{geometry} % Required for adjusting page dimensions

\usepackage[utf8]{inputenc}
\usepackage[T1]{fontenc}

\usepackage{fancyhdr}
\usepackage{enumitem}
\usepackage{lscape,multicol}
\usepackage{latexsym}
\usepackage{eurosym}

\usepackage{bm}
\usepackage{secdot}		
\usepackage{mathptmx}
\usepackage{float}
\usepackage[english]{babel}
\usepackage{textcomp}
\usepackage{tcolorbox}

\usepackage{xcolor} % ou xcolor selon l'installation
\definecolor{darkblue}{RGB}{0,99,124}
\definecolor{orange}{RGB}{255,88,0}
\definecolor{green}{RGB}{0,153,139}
\definecolor{blue}{RGB}{0,132,174}

\usepackage{mdframed}
\usepackage{multirow} %% Pour mettre un texte sur plusieurs rangées
\usepackage{multicol} %% Pour mettre un texte sur plusieurs colonnes
\usepackage{tikz}
\usepackage{url}
\usetikzlibrary{mindmap}

\usepackage{graphicx}
\usepackage[absolute]{textpos} 
\usepackage{colortbl}
\usepackage{array}
\newcolumntype{M}[1]{>{\raggedright}m{#1}}

\usepackage[T1]{fontenc} % Output font encoding for international characters
\usepackage[utf8]{inputenc} % Required for inputting international characters
\usepackage{XCharter} % Use the XCharter font
\usepackage{fancyhdr} % Needed to define custom headers/footers

\fancypagestyle{firstpage}{ % Page style for the first page with the title
	\fancyhf{}
}

\newcommand{\authorstyle}[1]{{\large\usefont{OT1}{phv}{b}{n}\color{darkblue}#1}} % Authors style (Helvetica)

\newcommand{\institution}[1]{{\usefont{OT1}{phv}{m}{sl}\color{darkblue}#1}} % Institutions style (Helvetica)

\usepackage{titling} % Allows custom title configuration

\newcommand{\HorRule}{\color{green}\rule{\linewidth}{1pt}} % Defines the gold horizontal rule around the title

\pretitle{% Move the entire title section up 

\vspace*{-2cm}
	\HorRule\vspace{0.5cm} % Horizontal rule before the title
	\includegraphics[height=1cm]{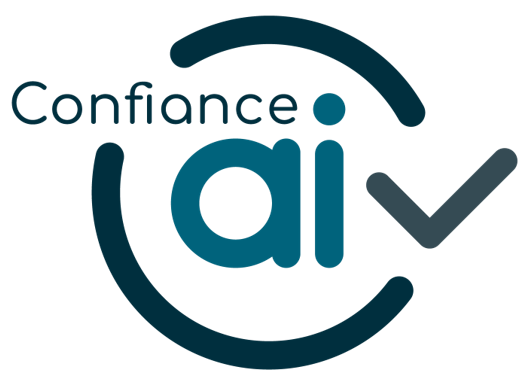}\\
	\fontsize{30}{36}\usefont{OT1}{phv}{b}{n}\selectfont % Helvetica
	\color{blue} % Text colour for the title and author(s)
}

\posttitle{\par\vskip 15pt} % Whitespace under the title

\preauthor{} % Anything that will appear before \author is printed

\postauthor{ % Anything that will appear after \author is printed
	\vspace{10pt} % Space before the rule
	\par\HorRule % Horizontal rule after the title
	\vspace{10pt} % Space after the title section
}

\usepackage{lettrine} % Package to accentuate the first letter of the text (lettrine)
\usepackage{fix-cm}	% Fixes the height of the lettrine

\newcommand{\initial}[1]{ % Defines the command and style for the lettrine
	\lettrine[lines=3,findent=4pt,nindent=0pt]{% Lettrine takes up 3 lines, the text to the right of it is indented 4pt and further indenting of lines 2+ is stopped
		\color{blue}% Lettrine colour
		{#1}% The letter
	}{}%
}
\usepackage{xstring} % Required for string manipulation
\newcommand{\lettrineabstract}[1]{
	\StrLeft{#1}{1}[\firstletter] % Capture the first letter of the abstract for the lettrine
	\initial{\firstletter}\textbf{\StrGobbleLeft{#1}{1}} % Print the abstract with the first letter as a lettrine and the rest in bold
}
\date{} 

\title{Empowering the trustworthiness \\ of  ML-based critical systems \\ through engineering activities} % The article title

\author{
	\authorstyle{Juliette MATTIOLI\textsuperscript{1}, Agnès DELABORDE\textsuperscript{2,3}, Souhaiel KHALFAOUI  $^{4,3}$, Freddy LECUE$^1$, Henri SOHIER$^3$ and Frédéric JURIE$^{5,3}$} % Authors
	\newline\newline % Space before institutions
	\textsuperscript{1}\institution{Thales, France}\\ 
	\textsuperscript{2}\institution{{L}aboratoire National de métrologie et d'Essais LNE }
	\\
	\textsuperscript{3}\institution{IRT SystemX, France}\\
	\textsuperscript{4}\institution{Valeo, France}\\
	\textsuperscript{5}\institution{Caen University, France}\\
}

\begin{document}

\maketitle % Print the title

\thispagestyle{firstpage} % Apply the page style for the first page (no headers and footers)

%------------------------
%	ABSTRACT
%------------------------

\lettrineabstract{This paper reviews the entire engineering process of trustworthy Machine Learning (ML) algorithms designed to equip critical systems with advanced analytics and decision functions. We start from the fundamental principles of ML and describe the core elements conditioning its trust, particularly through its design: namely domain specification, data engineering, design of the ML algorithms, their implementation, evaluation and deployment. The latter components are organized in an unique framework for the design of trusted ML systems.}

%%%%%%%%%%%%%%%%%%%%%%%%%%%%%%%%%%%%%%%%%%%%%%%%%
\section{Introduction}

Machine learning (ML) models are becoming inevitable components of Artificial Intelligence (AI) systems, including systems that require safety-critical environmental perception and decision-making. ML engineering \cite{treveil2020introducing,serban2021practices} is a new field leading to new issues and forcing companies to adapt their engineering practices and processes: 1) classic considerations on specification, traceability and validation are deeply challenged~\cite{bosch2021engineering,ozkaya2020really}; 2) processing data in ML algorithms requires new processes with new best practices~\cite{zinkevich2017rules}, as highlighted by ML Model Operationalization Management (MLOps) approaches; 3) advanced perception and complex decisions of a ML system must present new assesses trustworthiness through security, privacy, safety, explainability, etc. and other attributes related to specific concerns such as application domain concerns. To maximize the trustworthiness of ML-based critical systems, such attributes -- and the methods for concretely assessing their values -- must be clearly identified and mapped onto the ML processes and its lifecycle.

This paper presents ML algorithm engineering as a pipeline of processes and details the main challenges for reaching trustworthiness in the development procedures of an industrialized ML component. This paper is a result of the first year of the Confiance.ai program~\cite{braunschweig2022wall,chiaroni2021franco} which gathers multiple companies and research centers from different industries working on the development of trustworthy safety- and business-critical systems at scale.

%----------------
\section{ML Algorithm Engineering}

Algorithm engineering (AE) refers to the process required to bridge the gap between algorithm and concrete implementation in a dedicated programming paradigm to yield efficient, easily usable and well-tested implementations. This process encompasses a number of topics such as algorithm design, theoretical analysis, implementation, tuning, debugging evaluation, validation, and testing. Thus, AE is a methodology that combines theory with implementation and validation conducted through experimentation.

\begin{figure}[hbtp]
    \vspace*{-0.2cm}    
    \centering
    \includegraphics[width=\columnwidth]{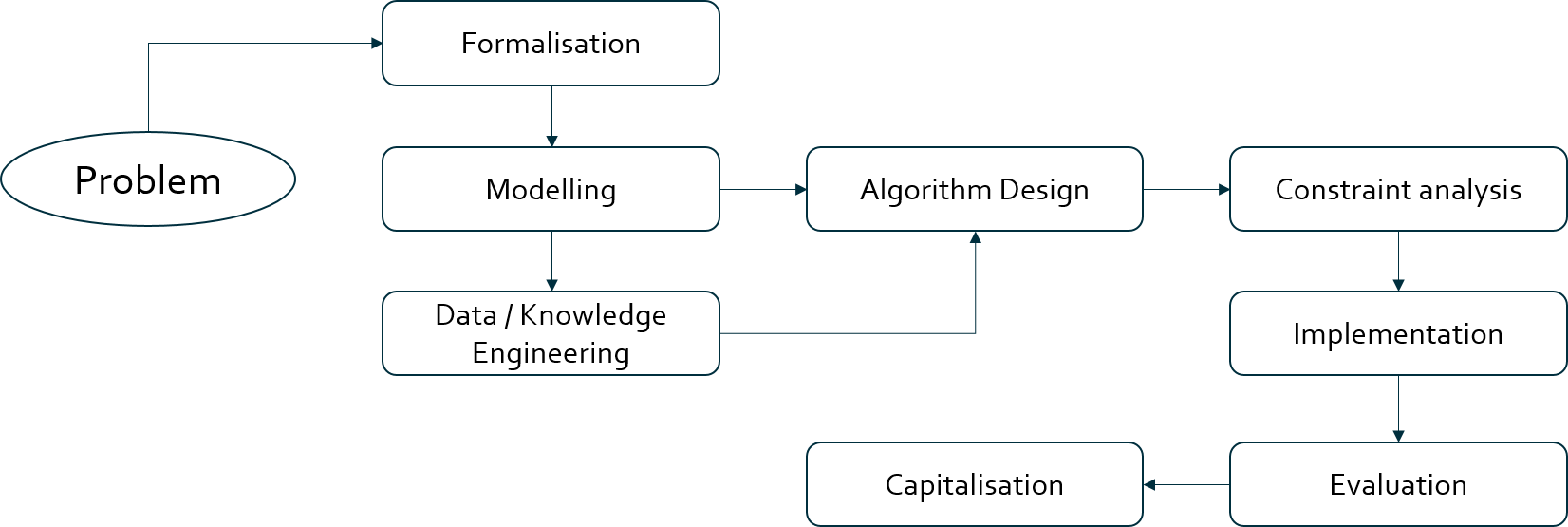}
\vspace*{-0.5cm}    \caption{AE process focusing on problem scoping, algorithm design and delivery}
    \label{fig:AE-lifecycle}
\vspace*{-0.3cm}   
\end{figure}

In essence, MLOps entails a set of practices and tools focused on software and systems engineering with close collaboration between ML developers, operation and engineering teams to improve quality of services while ML Engineering is often portrayed as the creation of a ML model, its fine-tuning, validation and deployment.

\begin{figure*}[hbtp]
\centering
 \vspace*{-1cm}    \includegraphics[angle=90,height=22.5cm]{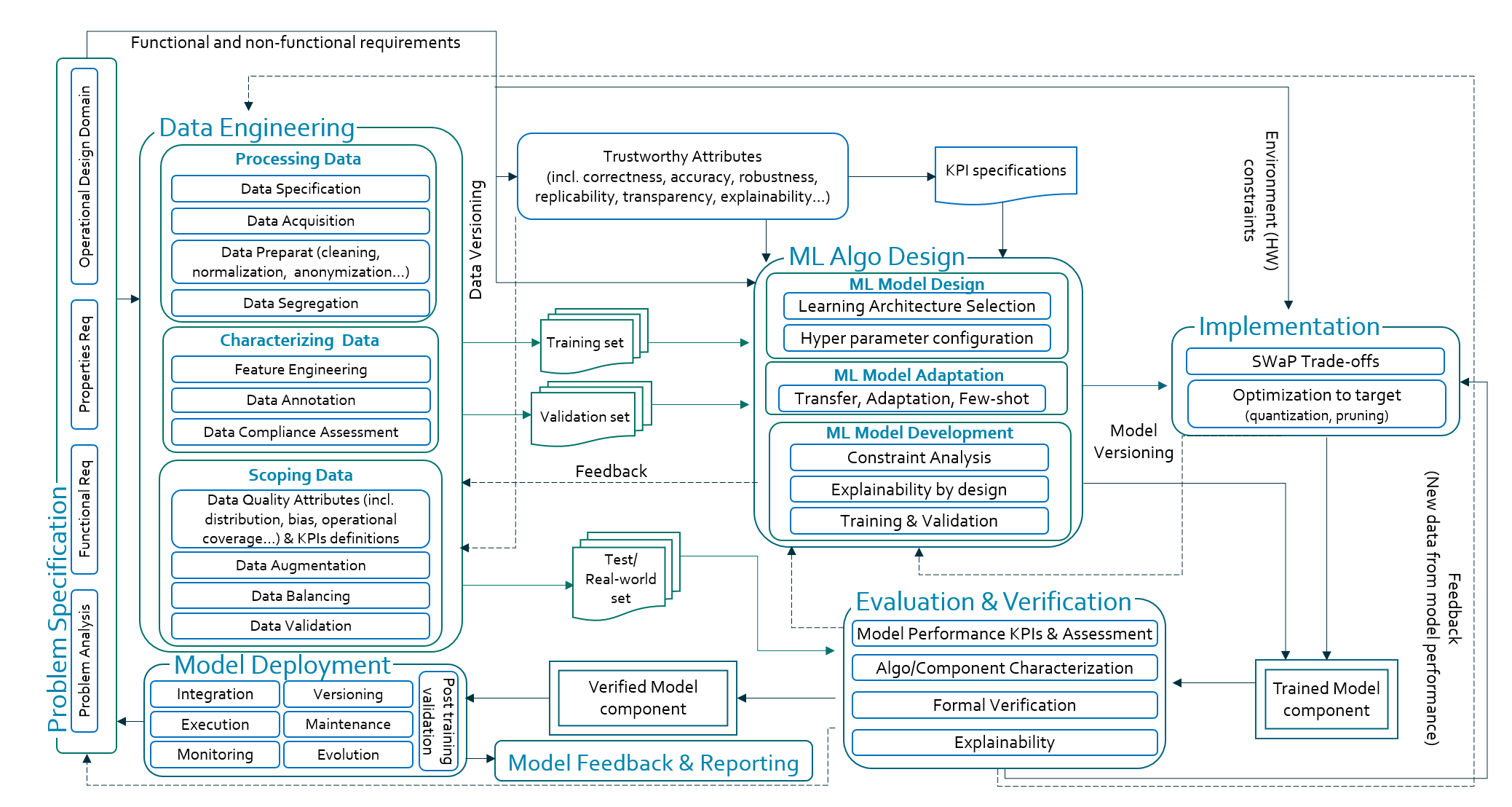}
 \vspace*{-0.75cm} \caption{Machine Learning Algorithm Engineering Pipeline}
   \label{fig:ML_pipeline}
\end{figure*}

In real-world industrial settings, the ML model is only a small part of the overall system and significant additional engineering and system functionalities are required to ensure that the ML model can operate in a reliable, predictable and scalable way with proper engineering of data and model pipelines, monitoring and logging, etc. To capture such issues, we present a ML algorithm engineering pipeline  (see fig.~\ref{fig:ML_pipeline}), where we distinguish requirements-driven development, safety-driven development  and ML-driven development. At the starting point, data must be available in size but also minimally prepared, validated and representative of the task at hand for training.

Main subtasks are encapsulated as a series of steps within the pipeline such as:
\begin{itemize}[noitemsep]
    \item \textbf{Problem specification}: It results in functional and non-functional requirements covering every aspect of the ML item: safety, performance, Operational Design Domain (ODD), etc. The ODD is the description of the specific operating condition(s) in which a safety-critical function or system is designed to properly operate as expected, including but not limited to environmental conditions and other domain constraints \cite{koopman2019many}. This usually drives the data collection task. 

    \item \textbf{Data engineering}: A ML model requires large amounts of data, which helps the model learn how to perform its purpose. Before it can be used, data need to be cleaned, organized, analyzed and visualized to support feature engineering. Data acquisition is the process of aggregating data into a homogeneous set. Among other properties, the collected data need to be sizable, accessible, understandable, relevant, reliable, and usable. Data preparation, or data processing, is the process of transforming raw data to make it usable for the model's purpose. Thus, one of the key challenges is to establish data sets that are of sufficient quality for training and inference. The importance of this task is highlighted by the data-first ML movement \cite{DBLP:journals/mlst/ZhouZJLN21}.

    \item \textbf{ML Algorithm Design}: The ML algorithm has to be designed or selected from existing ML libraries. By feeding a training set to the ML algorithm, it can learn appropriate parameters and features. Once training is complete, the model will be refined by using the validation dataset. This may involve modifying or discarding variables and includes a process of tweaking model-specific settings (hyperparameters) until an acceptable accuracy level is reached. Different models can be employed, validated and fine-tuned.

    \item \textbf{Implementation}: To develop a ML component, one has to decide on the targeted hardware and system platform, the IDE (Integrated Development Environment) and the language for development. These choices can typically impact the time-behavior or the power consumption. Embedded systems can be highly constrained.

    \item \textbf{Evaluation and verification}: Finally, after an acceptable set of hyperparameters is found and the model accuracy is optimized, the model is tested on a data set and/or assessed through formal verification. The evaluation can go beyond functional performance (such as accuracy) and encompass metrics relative to any other expected performance criteria. Example of such metrics are: explainability, interpretability, biais. Based on the feedback, one may return to training the model to adjust performance through output settings, or deploy the model as needed.
   
    \item \textbf{Model Deployment}: The integration of the ML model as a component in the overall system requires a tuning to the system characteristics, or indirectly to the environment of deployment; this adaptation can imply additional iterations of specification, development and testing. In addition, one should ensure that the model, once deployed, is monitored, that maintenance tasks can be performed, and that the model can be adapted to the evolution of the environment of deployment.

\end{itemize}

It is crucial to consider the evolution of the data, and its model for continuous deployment, as data might change overnight, then impacting the performance of models. Continuous integration, development, deployment and testing need to be always in the loop for MLOps.

The following paragraphs further detail the important subtasks of the ML algorithm engineering pipeline. However, we do not describe in more details the implementation of the models nor their deployments, as these tasks are very dependent on the computing targets.
%----------------
\section{Problem specification}

The first step of ML algorithm engineering process is to state the problem precisely. Operational requirements at the system level (i.e. where the system is considered as a black box) are derived into functional and physical requirements at the ML component level. The resulting requirements can for example be related to the component's functional performance, safety, integrability or maintainability. They can also for example combine customer requirements, operational constraints, regulatory restrictions, or implementation realities. In all cases, the functions should be evaluated for their safety related attributes.

Ideally, the design of a component should be based on a clear input domain and a clear input-to-output relation. However, in ML, the traditional programming paradigm is no longer suitable: 1) the environment's complexity is often difficult to reduce to a clear input domain; 2) instead of hard coding a clear input-to-output relation, one provides examples of inputs and outputs to a machine to generate the algorithms.

The requirements have to be refined and completed up to the point where they allow the development of the ML-based component. In particular, in the case of supervised machine learning, data could be considered as detailed requirements of the intended behavior of the ML-based component. Similarly, the structure of the ML model, its parameters and hyper-parameters could also be considered as detailed requirements for the ML-based component. The idea of the ODD \cite{gyllenhammar2020towards} can be used to indicate where ML-based critical systems can operate safely. In the automotive domain, an example of ODD are the closed roads, weather conditions, and presence of pedestrians or animals, etc. 

Furthermore, ML Model requirements should  express the expected properties of the ML Model with their acceptable tolerances. The ML Model specification activity steps are:
\begin{itemize}[noitemsep]
    \item ML Data Requirements are developed from the analysis of the subsystem requirements (including the ODD).
    \item ML Model requirements that specify nonfunctional requirements like performance objectives are stated in quantitative terms with tolerances where applicable.
    \item ML model requirements should define the ML Model outputs properties (e.g. boundaries, normalization). 
    \item ML Model requirements should define the expected ML Model response to robustness or generalization issues (e.g. specification of adversarial attacks) .
    \item Derived ML Model Requirements and the reason for their existence are defined.
    \item Derived ML Model Requirements, if any, are provided to the system processes, including the system safety assessment process.
    \item Each subsystem requirement allocated to ML Item should be covered by either ML Model requirement(s) or ML Data requirement(s) or both.
    \item ML Model requirements should be consistent with ML Data requirements.
    \item ML Data Requirements conform to the Requirements Standards, should be verifiable and consistent.
\end{itemize}
%----------------
\section{Data Engineering}
Data Engineering (DE) is a discipline that aims to organize, structure, trace and select data in such a way that its quality, availability, relevance and traceability can be guaranteed throughout the life cycle of the data. DE is then grouping all the engineering aspects of systems, processing, models and management of data, including but not limited to big data. \cite{ISO-25024:2015}, in the SQuaRE series of normative references for system and software quality, sets requirements and methods for the evaluation of the quality of data. In particular, this SQuaRE standard highlights the need that quality characteristics be “specified, measured, and evaluated whenever possible using validated or widely accepted measures and measurement methods”. DE is known that running ML end-to-end requires a large amount of time dedicated to preparing data, which includes acquiring, cleaning, organizing, analyzing, visualizing, and feature engineering. 

The goal of data acquisition is to find data sets that can be used to train ML models. This activity faces the following issue: how to search valuable data for the downstream task? 

After gathering the data from relevant sources we need to move forward to data engineering which aims at improving the quality of a data set. This stage helps us gain a better understanding of the data and prepares it for further evaluation. 

Data processing is the cornerstone of ML, as it will shape the input where data is the raw facts and figures, which could be structured and unstructured and acquisition means acquiring data for the given task at hand. Specifically, data sets tend to be unbalanced, have a high degree of heterogeneity, lack labels, tend to drift over time, contain implicit dependencies and generally require vast amounts of pre-processing effort before they are usable. 

\subsection{ODD-based data set specification}

Data sets should sufficiently cover the input domain. To reach this objective, descriptive attributes are used to characterize each data sample. These attributes correspond to explicit and interpretable operating parameters associated with the complex input space. This could come from the system requirements where operational design domain is specified. Thus, a data set specification (DSS) specifies a group of data elements and the conditions under which this group is collected. A DSS can define the sequence in which data elements are included, whether they are mandatory, what verification rules should be employed and the characteristics of the collection (e.g. its scope). Then, DSS consists in mapping required diversity to fully cover the operational design domain. The mapping produces an exhaustive list of attributes, which correspond to the dimensions that will be explored and sampled to achieve the targeted diversity and completeness.

These attributes are linked to the expected scenarios and conditions as well as semantic intra-class variability. The data sets should also be highly representative and complete, particularly regarding the coverage of corner case inputs.

\subsection{Data Segregation}
The fundamental goal of a supervised ML system is to use an accurate model based on the quality of its pattern prediction for data that it has not been trained on. As such, existing labeled data is used as a proxy for future/unseen data, this data segregation task involves breaking processed data into three independent data sets —- train, validation, and test:
\begin{itemize}[noitemsep]
    \item \textbf{Training set} is used to initially train the algorithm and teach it how to process information. This set defines model classifications through parameters, establishing the behavior of the machine learning model.
    \item \textbf{Validation set} is used to tune some hyperparameters of a model (e.g. number of hidden layers, learning rate, number of neurons per layer for neural netwoks), to anticipate some learning issues (overfitting, underfitting, etc.) and to estimate the accuracy of the model. 
    \item \textbf{Test set} is used to assess the accuracy and performance of the models. This set is meant to expose any issues or mistraining in the model. 
\end{itemize}

\subsection{Data Characterization and scoping}
\cite{nazabal2020data} have recently proposed to look at data engineering problem looking through the prism of the Data Organization, Data Quality issues, and Feature Engineering reading grid. 

\begin{itemize}[noitemsep]
    \item \textbf{Data Organization}: The first issue we face is producing a representation of the data that is well suited to the task at hand. 
    The first step is to structure the raw data so that it can be read correctly (data parsing). In a second step, a basic exploration of the data produces metadata for all elements (data dictionary). Then, data from several sources are combined into one extended table (data integration). In the last step, data is transformed from the original desired raw format.
    \item \textbf{Data Quality}: 
    Any problems in the data should be diagnosed, repaired or even removed. 
    Data quality problems relate to the data itself but not to the underlying structure of the data obtained earlier, which must be qualified independently. Common data cleaning operations include normalizing the data (canonicalisation), resolving missing entries (missing data), correcting errors or abnormal values (anomalies).
     \item \textbf{Feature Engineering}: Once the data is organized and cleaned, the next question is how to represent it in forms suitable for processing. 
    Feature engineering specifically applies to ML algorithms. Its processes can be based on some transformations of the raw data or even dedicated to creating new features from the raw data. This process usually relies on expert knowledge and is domain specific.
\end{itemize}

ML is generating renewed interest in data quality~\cite{mattioli2022IQ}. One understands that data qualification is a broad topic, that encompasses both the data itself and its relation to the ML algorithm, as well as a qualification of all the processes involved in the creation of the data set. Nevertheless, there is no consensus on what comprises the data quality dimensions. 

For example, \cite{batini2016data} proposed a classification framework where dimensions are included in the same cluster according to the similarity of the characteristics they measure. They end up with eight categories named after their representative dimension:
\begin{itemize}[noitemsep]
    \item Accuracy, correctness, validity and precision focus on the adherence to a given reality of interest.
    \item Completeness, pertinence and relevance refer to the capability of representing all and only the relevant aspects of the reality of interest.
    \item Redundancy, minimality, compactness and conciseness refer to the capability of representing the aspects of the reality of interest with the minimal use of informative resources.
    \item Readability, comprehensibility, clarity and simplicity refer to ease of understanding and fruition of data by users.
    \item Accessibility and availability are related to the ability of the user to access information from his or her culture, physical status/functions and technologies available.
    \item Consistency, cohesion and coherence refer to the capability of data to comply without contradictions to all properties of the reality of interest, as specified in terms of integrity constraints, data edits, business rules and other formalisms.
    \item Usefulness, related to the advantage the user gains from the use of information.
    \item Trust, including believability, reliability and reputation, catching how much information derives from an authoritative source. The trust cluster encompasses also issues related to security.
\end{itemize}

%----------------
\section{ML Algorithm Design}
This phase requires model technique selection and application, model training, model hyperparameter setting and adjustment, model validation, ensemble model development and testing, algorithm selection, and model optimization. Thus, this phase decides first the model type, variant and, where applicable, the structure of the model to be produced in the Model Learning stage. The process of adaptation is called training, in which samples of input data are provided along with desired outcomes. The algorithm then optimally configures itself so that it can not only produce the desired outcome when presented with the training inputs, but can generalize to produce the desired outcome from new, previously unseen data. This training is the “learning” part of ML.

Numerous types of ML techniques are available, including multiple types of classification models (to identify the category that the input belongs to) and regression models (to predict a continuous-valued attribute) for supervised tasks, clustering models (to group similar items into sets) for unsupervised tasks, and reinforcement learning models (to provide an optimal set of actions).  

A common question is “\textit{Which ML architecture should I use?}”. 
The DEEL project establishes the following table~\cite{delseny2021white} which gives a  short summary of the most common ML techniques, and indicates their main applications. Each kind of ML technique will rely on one or several hypothesis function space(s), and one or several exploration algorithms (not listed in this document) to minimize a loss function on the training dataset.

\noindent {\small \begin{tabular}{|p{3.5cm}|p{4cm}|}
\hline
\textbf{Techniques} & \textbf{Applications}  \\\hline
\textbf{Linear models}: Linear \& logistic regressions, SVM & 
Classification, Regression \\\hline
\textbf{Neighbourhood models}: KNN, K means, Kernel density & Classification, Regression, Clustering, Density estimation \\ \hline
\textbf{Trees}: decision trees,  regression trees & Classification, Regression \\ \hline
\textbf{Graphical models}: Bayesian network, Conditional Random Fields & Classification, Density estimation \\ \hline
\textbf{Combination of models}: Random Forest, Adaboost, XGboost & Classification, Regression, Clustering, Density estimation \\ \hline
\textbf{Neural networks, Deep Learning} & Classification, Regression\\ \hline
\end{tabular}}

\medskip 

After choosing the model, among the various algorithms present, one needs to tune the hyper parameters of each model to achieve the desired performance.
\begin{itemize}[noitemsep]
    \item Select the right algorithm based on the learning objective and data requirements.
    \item Configure and tune hyperparameters for optimal performance and determine a method of iteration to attain the best hyperparameters.
    \item Identify the features that provide the best results.
    \item Determine whether model explainability or interpretability is required.
    \item Develop ensemble models for improved performance.
    \item Test different model versions for performance.
    \item Identify requirements for the model's operation and deployment.
\end{itemize}

The resulting model can then be evaluated to determine whether it meets the business and operational requirements.
%----------------
\section{Evaluation and Verification}
Ensuring that a safety-critical system will perform adequately in their intended operational environment is a mandatory part of overall system validation. Traditional software validation includes traceability from requirements to system level tests. However, the use of ML techniques frustrates this approach due to the use of training  data rather than a traditional design process. In addition, software validation should be based on tests that show a level of performance that is adapted to the criticality of the risks, and performed on a data set that is fully representative of the factors of influence of the model. As previously mentioned however, the specification of the functional characteristics of the model and of the environment of operation may lead to the multiplicity of the factors of influence, and a valid demonstration of the performance of the model would imply relying on testing data sets of huge volume (in the worst case, millions of sample). Verification through formal methods or by simulation are interesting tracks to fulfill this goal, but they are still at an early stage of research.

Verification therefore requires at least ensuring that training data and testing data cover all \textit{relevant} operational conditions. Making this problem tractable in practice is generally accomplished by constraining the operational environment to a subset of all possible situations that could be dealt with by a human operator. That approach to limiting the operational needs of the system is known as adopting an ODD \cite{koopman2019many}. 

The testing of an ML component aims at detecting gaps between achieved and intended (targeted) behaviors of ML models. Formally, ML testing refers to any activity designed to reveal ML bugs, where a ML bug refers to any imperfection in a ML item that causes a discordance between the output of the model and the output of reference. Examples of gaps could be due to shift in training and testing data distribution, or wrong assessment of data fit to the task at hand, therefore data is usually the cause of wrong or unexpected errors. 

This definition underlines three preliminary challenges to overcome. First, ML system may have different types of ‘required conditions’, i.e. properties that should be verified --  we may classify them into basic functional requirements (e.g. correctness and model relevance) and non-functional requirements (e.g. efficiency, robustness, fairness, interpretability). The verification of such properties requires the use of different methods and metrics, which means that the selection of the best tools for the verification of the component must be preceded by a definition of the required conditions: "What do we want to prove through testing?". Secondly, an ML bug may exist in the data, the learning program, or the framework. Here again, this means that the testing strategy should either address the component itself, or question other "sub-component", which may make the testing more complex since establishing a causal link between the bug and its source may be difficult, and the definition of a testing protocol allowing the distinction of independent and dependent variables is not trivial in an ML pipeline. Finally, the notion of testing activity may encompass several radically different approaches for testing. This may include test input generation, test oracle identification, test adequacy evaluation, and bug triage. The selection of the approach must be based on a trade-off between the technical feasibility of performing such test on the ML component and the required conditions initially formalized.

\subsection{Quality Control}

Quality control (QC) is an essential part of the verification and validation of the ML component. QC may be performed through an estimation of the success of the task solved by the component. Traditional metrics for regression problems include Mean Squared Error (MSE) or Mean Absolute Error (MAE), while classification problems can be evaluated through precision, accuracy and recall. In classification problems, a confusion matrix (depicting the distribution of true/false negatives/positives for each class) is a practical tool for  visualizing of the errors, and allows the computation of most metrics (precision, recall, sensitivity, specificity, F1 score, ROC curve, etc.).

The most common evaluation protocol consist in maintaining a hold-out validation set. This consists on setting apart some portion of the data as the test set. The process would be to train the model with the remaining fraction of the data, tuning its parameters with the validation set and finally evaluating its performance on the test set. The reason to split data in three parts is to avoid information leaks. The main inconvenient of this method is that if there is small amount of data available, the validation and test sets will contain so few samples that the tuning and evaluation processes of the model will not be effective. An alternative is k-Fold, which consists in splitting the data into k partitions of equal size. 

Another interesting approach is: Iterated k-fold validation with shuffling. This technique is relevant when having little data available and it is needed to evaluate models as precisely as possible. 

Functional performance evaluation has its own challenges. The selection of the most adapted metrics to reflect the desired level of performance, as well as the selection of a suitable protocol for testing, require careful work. However, the notion of QC should go further beyond a simple estimation of functional performance.

First, we note that the recourse to a validation set, in this context, is part of the ML algorithm design step. Here the focus is made on the technical validity of the algorithm design. This means that there are only few links between this testing activity and, for example, the operational constraints established in the specification phase. In the same idea, the influence of the training data is ignored at this stage, since traditional protocols do not necessarily take into account the informational value of the data points in each set (hence, a risk for representativeness, or a risk of ignoring corner case values either in training or testing). QC should then encompass more than a simple evaluation procedure of the ML algorithm: QC procedures should be formalized and deployed ideally at each stage of the ML pipeline, with different objectives and verification strategy for each stage, but with one overarching objective in my mind: "How can I ensure the quality of all the processes involved in the development of the ML component".

Although each domain has their own traditional ways of performing qualification (for example, data qualification has its own procedures), their link with the particularities and the constraints of ML components is not always well established. In addition, some aspects of the verification and validation strategies are underestimated, or at least not part of the routine, in ML engineering. For example, information emanating from data engineering about the limits and constraints of the data should reflect in the overall strategy of the evaluation of the model. The system in which the ML component is intended must also provide its own set of constraints with which to check the compliance of the component. 

This means that all brick of the ML pipeline should include specific QC procedures, and the information should propagate to the relevant bricks of the pipeline and condition the overall evaluation of the quality of the component.

\subsection{Algorithm / Component Characterization}

\begin{figure}[tb]
       \includegraphics[width=8.5cm]{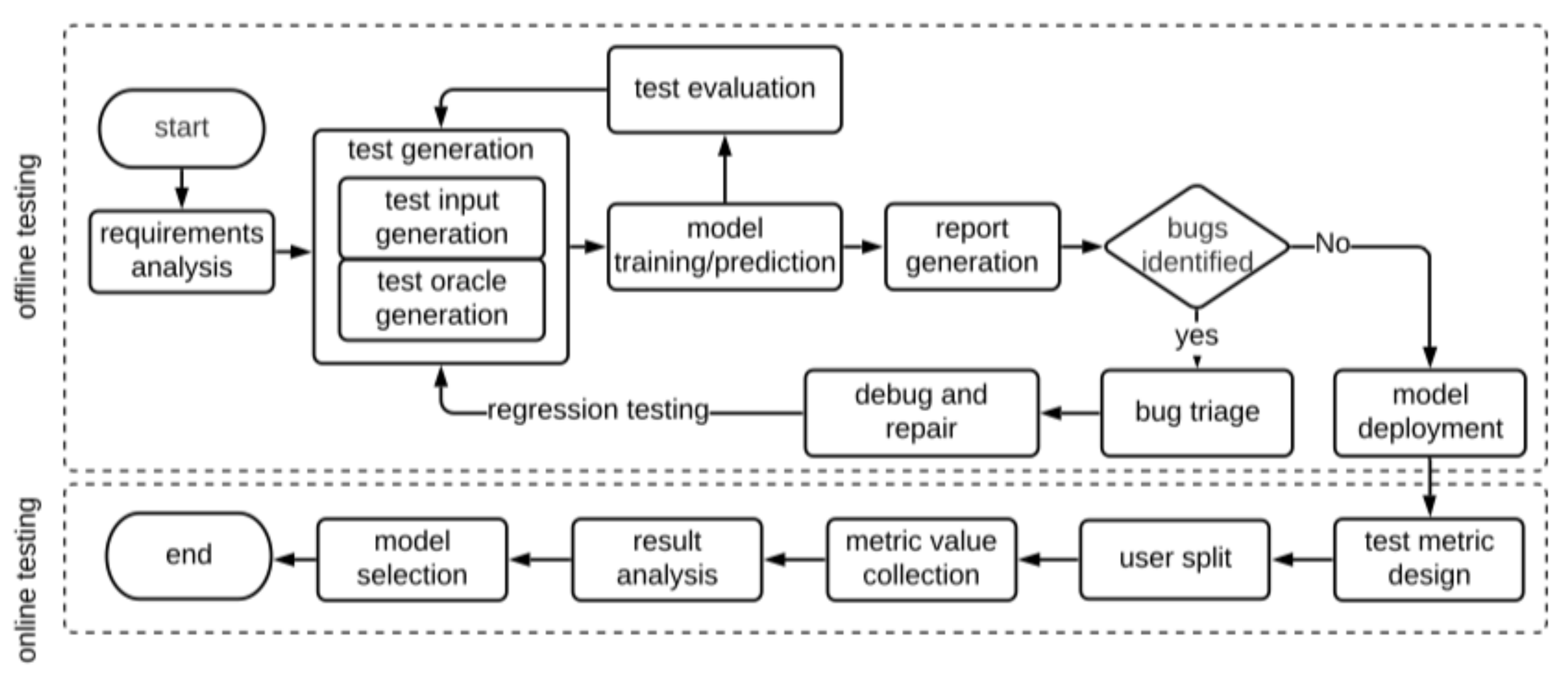}
    \vspace{-0.4cm}
    \caption{Idealized Workﬂow of ML testing}
    \vspace{-0.5cm}
    \label{fig:my_label}
\end{figure}

ML software includes many different components (data, learning algorithms, etc.). The testing phase must therefore include steps to verify each ones of these components. 

The behaviors of ML-based systems depend on the data used to train them. Any problems in the data affect the quality of the resulting model, and by cascade effect, may lead to other problems in the operation of the model. Consideration should be given to the ability of the data to train or evaluate a particular model (completeness of the data), whether the data is representative of the data that the system will have to process, whether the data contains a lot of noise (particularly on labels), whether there is a bias between training and test data, whether there is any poisoning of the data or data containing adversarial noise. There are also different methodologies that can be used to search for such bugs.

ML, especially the phase for training the model, is computationally intensive. Deep learning frameworks (e.g., pytorch, tensorflow, etc.) simplify the writing of learning programs, making it easier for developers. Therefore, they play a more important role in ML development than in traditional software development. For this reason, it is important to test these frameworks and check that they do not contain any bugs. Various authors have shown that the most widely used frameworks are far from being bug-free, and have proposed methodologies for testing these frameworks: \cite{xiao2018security18,guo2018orchestrated18,sun2017empirical}. 

It is also necessary to detect bugs that may occur in the model training software. These software have two components: the optimization algorithm designed by the developer or taken from a framework, and the actual training code that developers write to use, deploy or conﬁgure the optimization algorithm. A bug in the training phase can come from the design of the optimizer, a misconfiguration, or a misuse of the optimizer, as well as from errors in the code using it. We can cite, as an example, \cite{schaul2013unit}, who proposed unit tests designed for stochastic optimization. They can be used to test learning algorithms in order to detect bugs as early as possible. \cite{zhang2020machine} proposed an idealized workﬂow of ML testing, detailing the different component, which we reproduce in  Figure \ref{fig:my_label}.

\subsection{Certification and Assurance Case}
Certification standards should impose neither a specific model nor a specific training technique. The focus should rather be on the properties, such as explainability and  robustness that the model must possess after training. Other properties such as maintainability, auditability, etc. could also be checked at this stage. The depth of demonstration of these properties can vary depending on the requirements.  If these properties are required for the overall safety demonstration, then in-depth demonstration is necessary. To illustrate that purpose, the Federal Aviation Administration (FAA) launched in 2016 an initiative called “Overarching Properties”. The objective of this initiative is to develop a minimum set of properties such that if a product is shown to possess all these properties, then it can be certified. As of 2019, the three overarching properties retained are:
\begin{itemize}[noitemsep]
    \item Intent. The defined intended functions are correct and complete with respect to the desired system behavior.
    \item Correctness. The implementation is correct with respect to its defined intended functions, under foreseeable operating conditions.
    \item Innocuousness. Any part of the implementation that is not required by the defined intended behavior has no unacceptable safety impact.
\end{itemize}

These properties are, by construction, too abstract to constitute an actionable and complete means of compliance for certification. In practice, they shall be refined to be applicable, leaving an opportunity to establish a specific set of methods for the implementation and verification of these properties for the certification of ML systems.

However, following the FAA Initiative, if no requirement stems from the safety assessment and component specification, then the ML model could remain a “black box”, without explainability and/or robustness demonstration. Some verification activities can be performed directly on the model, before implementation.  If it is the case, it should be demonstrated that the results of these verification activities are preserved after implementation.

The assurance of a system is typically communicated in the form of an assurance case, capturing “\textit{a reasoned and compelling argument, supported by a body of evidence, that a system, service or organization will operate as intended for a defined application in a defined environment}”. 

\subsection{Explainability}
A first way to assess the explainability of results produced by ML is through a human evaluation. This is, to date, the most reliable approach to assessing explainability. \cite{doshi2017towards} have proposed a taxonomy of explainability assessment methods. This human-based approach uses the results of human evaluation on simpliﬁed tasks. The second one, based on function, does not require human experiments but uses a quantitative metric as a proxy for the quality of the explanation (e.g. through the depth of a decision tree).

Automatic assessment of explainability allows for more objective procedures and easier scalability than human assessment. However, it requires the definition of metrics, which are not easy to establish and may depend on the application domains. As an illustration, we can cite \cite{cheng2018towards} who analyzed the impact of object masking in the image domain, \cite{zhou2018metamorphic} deﬁned the concepts of metamorphic relationship models useful to help end-users understand the operation of an ML system.
%----------------
\section{Conclusion}
As any critical system, a critical system which embeds ML needs to have well defined development methods from its design to its deployment and qualification. This requires a complete tool chain ensuring trust at all stages, as: 
\begin{enumerate}[noitemsep]
    \item Specification, knowledge and data management; 
    \item Algorithm and system architecture design;     
    \item Characterization, verification and validation of ML functions;  
    \item Deployment, particularly on embedded architecture;     
    \item Qualification, certification from a system perspective.
\end{enumerate}

To guarantee a trustworthy algorithmic design (robust, reliable...), we presented how algorithm engineering can integrate the ML paradigms and specific challenges that arise. We noted that the ML engineering pipeline requires several specific activities meant to ensure the overall trustworthiness, in terms of design and evaluation. Depending on the stage of the ML engineering cycle, several properties should be assessed, and the approaches should leverage knowledge and best practices for various disciplines.

In addition, the safety and security of critical systems which embed ML require the demonstration of the following four properties:
\begin{itemize}[noitemsep]
    \item Validity: to guarantee that an AI-based system will do what it is meant to do -- everything that it is meant to do and just what it is meant to do.
    \item Security: to ensure robustness and resilience to adversarial conditions, such as decoying and cyberattacks.
    \item Explainability: to be able to provide human-level, understandable and context-relevant justifications and explanations.
    \item Responsibility: to be compliant with ethical, legal and regulatory frameworks.
\end{itemize}

Here, the robustness characterizes its ability to provide correct answers in the face of unknown situations or maliciousness. However, this property is harder to prove than accuracy. Indeed, a non-accurate system cannot be robust. But more importantly, an accurate system may not be robust. This is the case of a learning-based system that has memorized the training data and will make wrong decisions in the future based on new data. This phenomenon is called \textit{overfitting}.

Moreover, ML remains vulnerable, and if one is not careful, particularly sensitive to so-called "adversarial" attacks, attacks that take advantage of the functioning of the underlying algorithms to generate small perturbations in the analyzed data and force the AI to return an incorrect result. Many defenses have been proposed in the last few years by the scientific community but are sometimes refuted with new attacks making them obsolete. This is why it is necessary to develop methods and tools to design robust algorithms and at least characterize their robustness.

It is also necessary to prove that ML-based critical systems are controllable, i.e.  well-founded or consistent, if it can be proved that they only do what is expected of them. The questions related to the problems of robustness and consistency are beginning to be the subject of work related to formal proofs. The latter aim at providing a priori guarantees on the reliability of a system, contrary to validation methodologies by direct experimentation which aim at providing a posteriori guarantees. Finally, understanding AI and its reasoning is necessary to determine how much we can trust it. 

For ML approaches, data are therefore crucial for learning, testing and validation. It is not enough to have a lot of data, it must be of "good quality" and representative of the domain of use of the system concerned, without which these approaches give poor results. New methodologies need to be defined for a better control of data acquisition, exploration, enrichment, annotation and preparation stages.

An algorithm engineering based approach for ML-based safety critical system allows a sound definition of each separate steps to conduct the development of an ML component. This allows in particular the identification of the variety of tasks, activities and fields of expertise involved in the development, and helps spotting the properties of trustworthiness that need to be checked. The development of a trustworthy ML component still requires an important amount of research and practice towards a comprehensive framework providing all the necessary guarantees of compliance.

\subsection*{Acknowledgment}  
This work has been supported by the French government under the "France 2030” program, as part of the SystemX Technological Research Institute Research Institute

%---------------- Biblio

\bibliographystyle{apalike}
\bibliography{ijcai22}

%----------------
\end{document}